\documentclass[namedreferences]{solarphysics}
\pdfoutput = 1
\usepackage[hyperref,optionalrh,solaromanenum]{spr-sola-addons} 
\usepackage{graphicx}                    
\usepackage{color}                       
\usepackage{url}                         
\usepackage{dirtytalk}                   
\usepackage{graphicx}
\usepackage{subfig}

\newcommand*{\unit}[1]{\ensuremath{\mathrm{\,#1}}} 

\begin{document}

\begin{article}

\begin{opening}

\title{Sky Brightness evaluation at Concordia Station - Dome C, Antarctica for ground-based observations of the Solar Corona}


\author[addressref={aff1, aff2}, corref, email={alessandro.liberatore@inaf.it}]{\inits{A.}\fnm{Alessandro}~\lnm{Liberatore}}
\author[addressref=aff1, corref, email={gerardo.capobianco@inaf.it}]{\inits{G.}\fnm{Gerardo}~\lnm{Capobianco}}
\author[addressref=aff1, corref, email={silvano.fineschi@inaf.it}]{\inits{S.}\fnm{Silvano}~\lnm{Fineschi}}

\author[addressref=aff1]{\inits{G.}\fnm{Giuseppe}~\lnm{Massone}}
\author[addressref=aff1]{\inits{L.}\fnm{Luca}~\lnm{Zangrilli}}
\author[addressref=aff1]{\inits{R.}\fnm{Roberto}~\lnm{Susino}}
\author[addressref=aff1]{\inits{G.}\fnm{Gianalfredo}~\lnm{Nicolini}}


\runningauthor{Liberatore A. et al.}
\runningtitle{Sky brightness at Concordia Station, Antarctica}

\address[id=aff1]{INAF - National Institute for Astrophysics; \\
OATo -  Astrophysical Observatory of Turin; \\
Via Osservatorio 20, 10025, Pino Torinese (To), Italy.}
\address[id=aff2]{University of Torino, Physics Department \\
Via Pietro Giuria 1, 10125, Torino, Italy.}

\begin{abstract}
The evaluation of sky characteristics plays a fundamental role for many astrophysical experiments and ground-based observations. In solar physics, the main requirement for such observations is a very low sky brightness value, that is, less than $10^{-6}$ of the solar disc brightness ($\unit{B_\odot}$). Few places match such requirement for ground-based, out-of-eclipse coronagraphic measurements. One of these places is, for instance, the Mauna Loa Observatory ($\approx3400\unit{m}$ a.s.l.).
Another candidate coronagraphic site is the Dome C plateau in Antarctica. In this article we show the first results of the sky brightness measurements at Dome C with the Extreme Solar Coronagraphy Antarctic Program Experiment (ESCAPE) at the Italian-French Concordia Station, on Dome C, Antarctica ($\approx3300\unit{m}$ a.s.l.) during the summer 34th and 35th Expeditions of the Italian Piano Nazionale Ricerche Antartiche (PNRA).  The sky brightness measurements were carried out with the internally-occulted Antarctic coronagraph, AntarctiCor. 
In optimal atmospheric conditions, the sky brightness of Dome C has reached values of the order of $1.0 - 0.7 \times 10^{-6} \unit{B_\odot}$. 
\end{abstract}


\end{opening}


\section{Introduction} 
\label{s:introduction} 
The Sun has an atmosphere divided into several layers. The outermost is called solar corona. It consists of plasma at very high temperatures (up to $10^6$ K) that extends millions of kilometres into outer space. The solar corona is the results of three different main contributions: the K-Corona, due to Thomson scattering by the free coronal electrons of the photospheric radiation; the F-Corona, due to diffusion of solar radiation by dust particles and E-Corona due to emission processes by coronal ions identified as \say{forbidden} lines.

\begin{figure} 
    \centerline{\includegraphics[width = 0.78\textwidth, clip=]{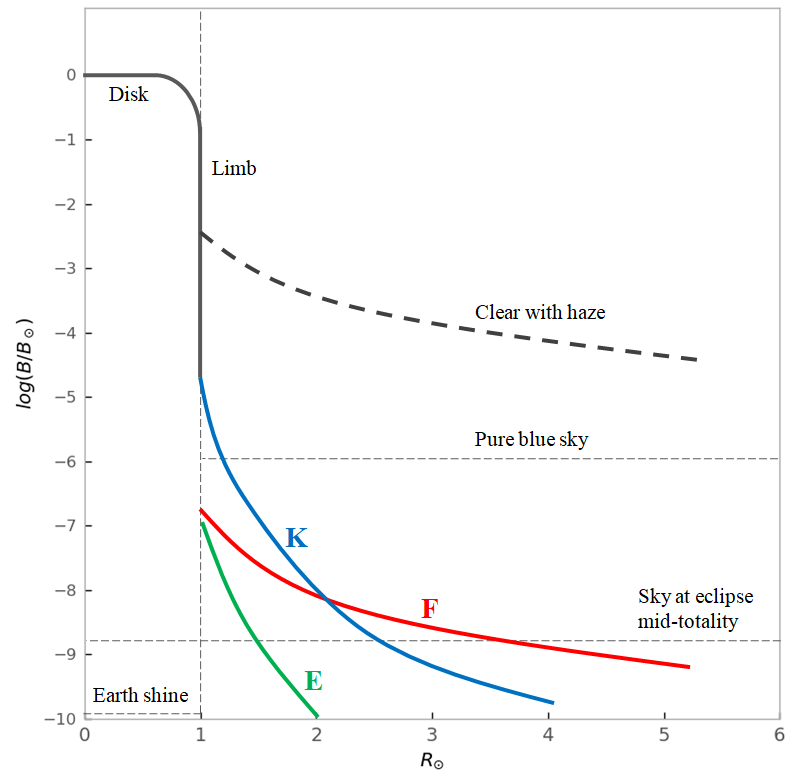}}
    \caption{Brightness of the different solar corona components ($B/B_\odot$) in function of the heliocentric distance. At least a $B_{sky}[B_{\odot}]\approx10^6$ is necessary to have a coronagraphic sky.}
    \label{fig:brightness_solar_corona_components}
\end{figure}

The brightness of each component decreases with a power law moving away from the Sun~\citep{Phillips_1992_SunGuide, Laurence_1996}. Figure~\ref{fig:brightness_solar_corona_components} shows that the brightness of the sky (\say{clear with haze}) makes ground-based observations of the solar corona difficult. Total solar eclipses give the opportunity to observe the corona with a reduced sky brightness. However the short duration of these events (max 7.5 minutes), the possibility of adverse weather conditions and the frequent need of accessing remote locations for the observing sites, makes it difficult to carry out continuous and detailed coronal studies. A \say{pure blue sky} for ground-based coronagraphic observations is defined as $\approx10^{-6}$ of the Sun's disc brightness ($\unit{B/B_\odot}$). Ground-based observations of the solar corona were made possible with the development of the internally-occulted coronagraph by Lyot~\citep{Lyot_1932}.

Mauna Loa, in the Big Island of Hawai'i is a coronagraphic site hosting the Mauna Loa Solar Observatory (MLSO) operated by the US High Altitude Observatory. MLSO can carry out systematic observations of the solar corona thanks to a sky-brightness value of $\approx1-5\times10^{-6}\unit{B_\odot}$ for different wavelengths~\citep{MaunLoa_skyB_2015}. One of the goals of the \textbf{E}xtreme \textbf{S}olar \textbf{C}oronagraphy \textbf{A}ntarctic \textbf{P}rogram \textbf{E}xperiment (ESCAPE:~Fineschi et al.,~\citeyear{Fineschi_2019_ESCAPE}) is the determination in Antarctica of a location with a \say{coronagraphic sky} that would allow systematic ground-based observations of the solar corona. In the next section we will introduce the ESCAPE project instrumentation and goals (Sec.~\ref{s:ESCAPEProject}) and we will show the results of sky-brightness measuraments (Sec.~\ref{s:Results}) during the 34th (austral summer 2018/2019) and 35th (austral summer 2019/2020) Italian Expeditions in Antarctica, Concordia Station - Dome C plateau at $\approx3 233\unit{m}$ above sea level.

\section{ESCAPE Project} 
\label{s:ESCAPEProject} 
\begin{figure}[ht]
    \centerline{\includegraphics[width = 1.00\textwidth, clip=]{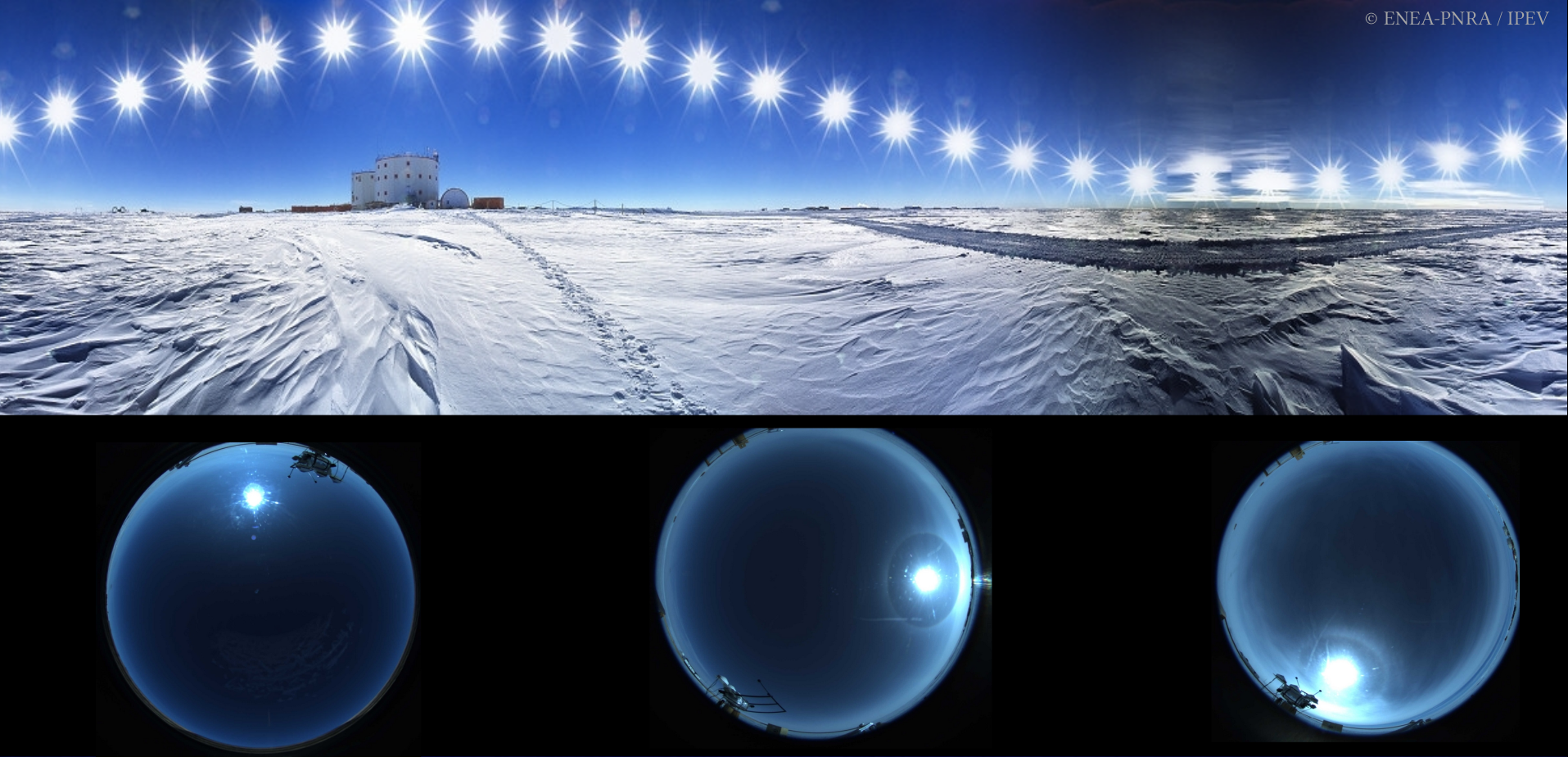}}
    \caption{Composition of hourly images showing the Sun position at Concordia Base during the Antarctic summer (Credits: Guillaume Dargaud). In the bottom, 360\textdegree~images from Baseline Surface Radiation Network (BSRN) project - PI, Dr. A. Lupi~\citep{lupi2021baom}. It is possible to observe a very good sky condition on the left image. In the middle, the presence of a solar halo is a sign of ice crystals in the atmosphere that can potentially compromise observations by producing straylight. Worst sky condition are shown on the right with a slight cloud cover as well. By using the full sky camera it is possible to have a general view of sky conditions to avoid the worst days.}
    \label{fig:sun24h}
\end{figure}

Antarctica offers a great opportunity for ground-based observations of the solar corona. Actually, the high altitude of the Antarctic plateau of Dome~C ($\approx3 233\unit{m}$ a.s.l.), the high latitude (75\textdegree~06$^\prime$~S; 123\textdegree~20$^\prime$~E), the large amount of daily hours of observations during the Antarctic summer (Fig.~\ref{fig:sun24h}) and the almost total absence of anthropic pollution, are necessary conditions for a low sky-brightness. 

First attempts to characterise the sky-brightness at Dome C were perfomed in 2008 by the pioneering observations of J. Arnaud~\citep{Faurobert_Arnaud_2012}.
One of the goals of the ESCAPE Project, at the Italian-French Station Concordia, is to quantitatively evaluate the sky brightness at the Dome C plateau. Within the ESCAPE project we developed an internally-occulted antarctic coronagraph (AntarctiCor) for observations of the K-corona polarized brightness ($pB$) generated by Thomson scattering of photospheric light of coronal free electrons for the determination of the coronal electron density~\citep{HC_VanDeHulst_1950}. 

\begin{figure}[ht]
    \centerline{\includegraphics[width = 1.00\textwidth, clip=]{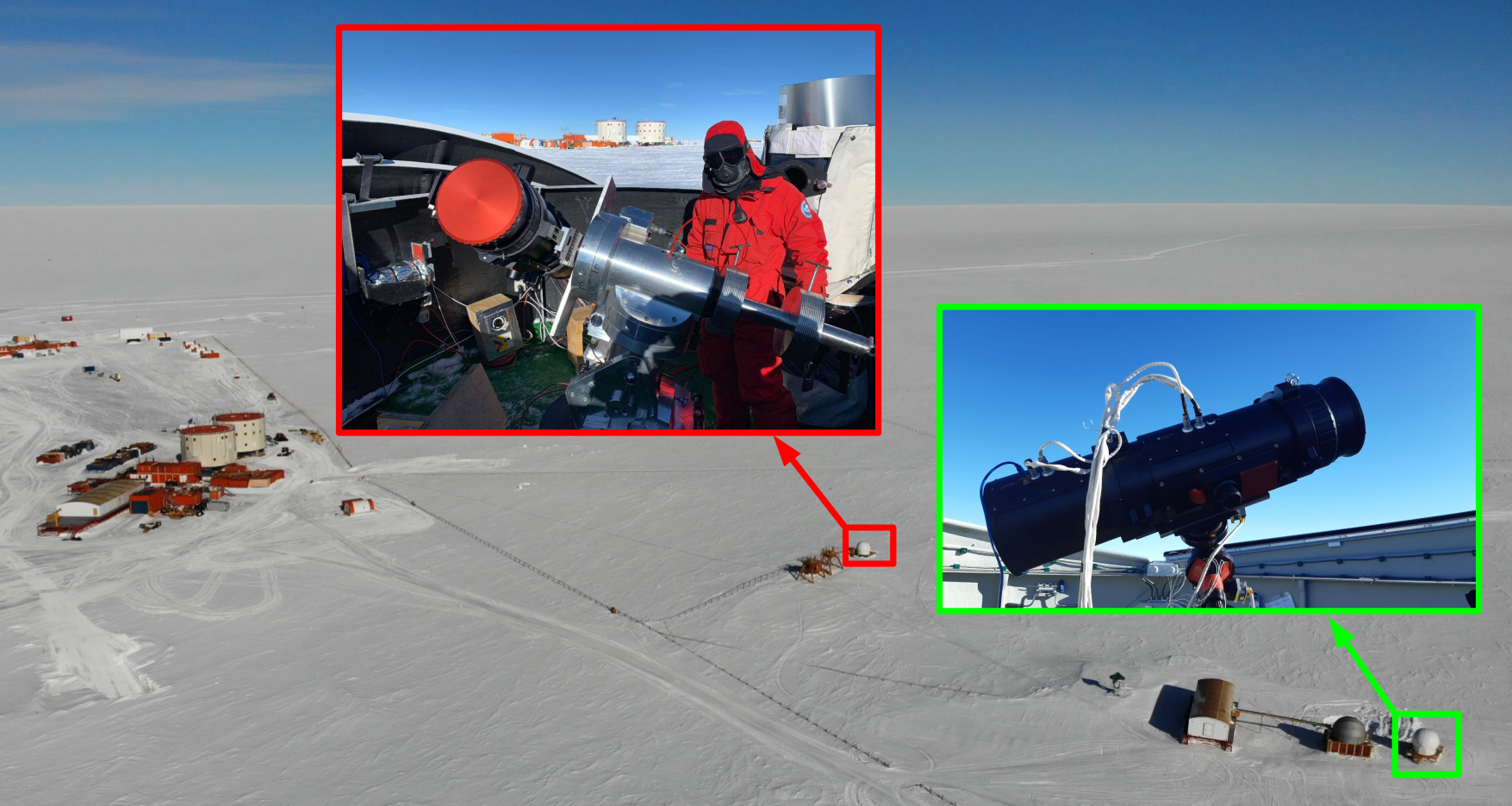}}
    \caption{AntarctiCor hosted by the Antarctic Search for Transiting ExoPlanet (ASTEP) project equatorial mount during the 34th Expedition (red box) and in the Baeder dome during the 35th Expedition (green box) at Concordia station for the ESCAPE project. (Credits: A. Liberatore and G. Capobianco  @PNRA/IPEV)}
    \label{fig:Montato}
\end{figure}

In the following section, we will report the sky-brightness measurements. The coronal images are still under analysis. More details about ESCAPE and its science objectives can be found in~\cite{Fineschi_2019_ESCAPE}. The evaluation of the sky-brightness was performed during  the 34th and 35th Italian Expeditions in Antarctica by using the Antarctica solar Coronagraph, AntarctiCor (Fig.~\ref{fig:Montato}).

\subsection{Antarctic coronagraph instrument (AntarctiCor)} 
\label{ss:AntarcticCoronagraph} 

\begin{figure} 
    \centerline{\includegraphics[width = 1.00\textwidth, clip=]{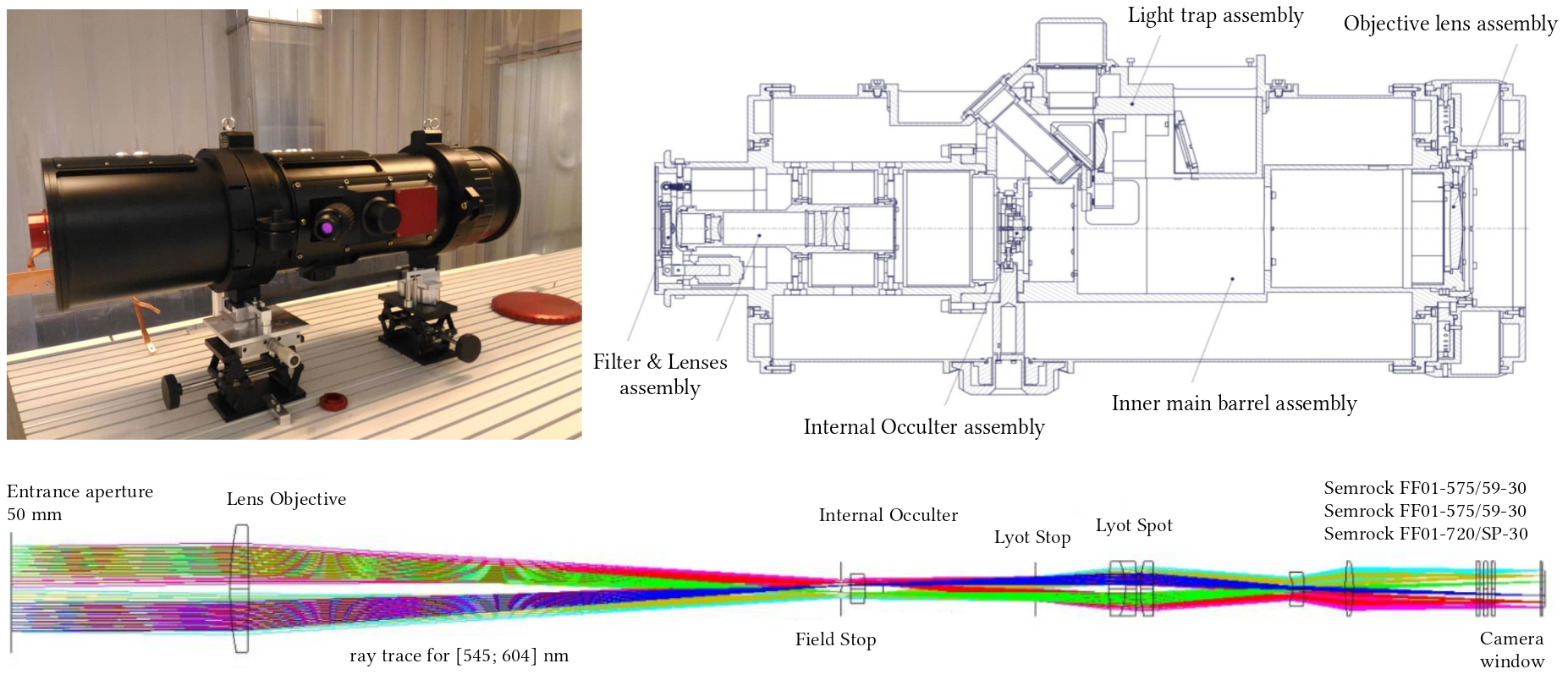}}
    \caption{\textbf{Top left:} AntarctiCor in the INAF Optical Payload Systems facility (OPSys) - clean room ISO 5 in Turin (Italy) for tests and calibrations~(\citeauthor{Fineschi_2019}~\citeyear{Fineschi_2019}). The main subassembly diagram (top roght panel) comprises: 1. objective lens assembly; 2. inner main barrel assembly; 3. internal occulter assembly; 4. lenses assembly; 5. filter assembly; 6. light trap assembly; 7. microscope assembly. \textbf{Bottom:} AntarctiCor ray tracing for the wide-band $(591 \pm 5)\unit{nm}$ optical path.}
    \label{fig:AntarctiCor}
\end{figure}

The instrument deployed during both campaigns was the Antarctica solar Coronagraph, AntarctiCor (Fig.~\ref{fig:AntarctiCor}). The main features of the instrument are summarized in Table~\ref{tab:AntarctiCor_features}. It is a classical Lyot internally-occulted coronagraph~\citep{Lyot_1932} based on the externally-occulted ASPIICS (Association de Satellites pour l'Imagerie et l'Interferométrie de la Couronne Solaire) coronagraph for the European Space Agency (ESA) formation-flying PROBA-3 (Project for On-Board Autonomy-3) mission~\citep{Galy_2015}.

\begin{table}[ht]
  \caption[example]{AntarctiCor Characteristics~(\citeauthor{Fineschi_2019}~\citeyear{Fineschi_2019}).} 
    \label{tab:AntarctiCor_features}
     \begin{tabular}{l|l}
        \hline
        \textbf{Telescope design} & Classical internally-occulted Lyot coronagraph\citep{Lyot_1932} \\
        \textbf{Aperture} & 50\unit{mm} \\
        \textbf{Eff. Focal Length} & 700\unit{mm} \\
        \textbf{f/ratio} & 14 \\
        \textbf{Spectral Ranges} & $(591 \pm 5)\unit{nm}$ -- see Fig.~\ref{fig:Filter}\\
        \textbf{Camera type} & Interline transfer CCD PolarCam; model: U4 \citep{Tech4D_PolarCam} \\
        \textbf{Camera format} & $1950 \times 1950$\unit{pixels} \\
        \textbf{Pixel size} & $7.4\unit{\mu m} \times 7.4\unit{\mu m}$ \\ 
        \textbf{Plate scale} & 4.3\unit{arcsec/pixel} \\
        \textbf{Field of View (FoV)} & $\pm0.6$\textdegree~$\equiv \pm2.24\unit{R_\odot}$ \\
        \textbf{Polarisation analysis} & Spatial modulation by linear micropolarisers on the sensor \\
        \hline
        \end{tabular}
\end{table} 

\begin{figure}[ht]
    \centerline{\includegraphics[width = 0.50\textwidth, clip=]{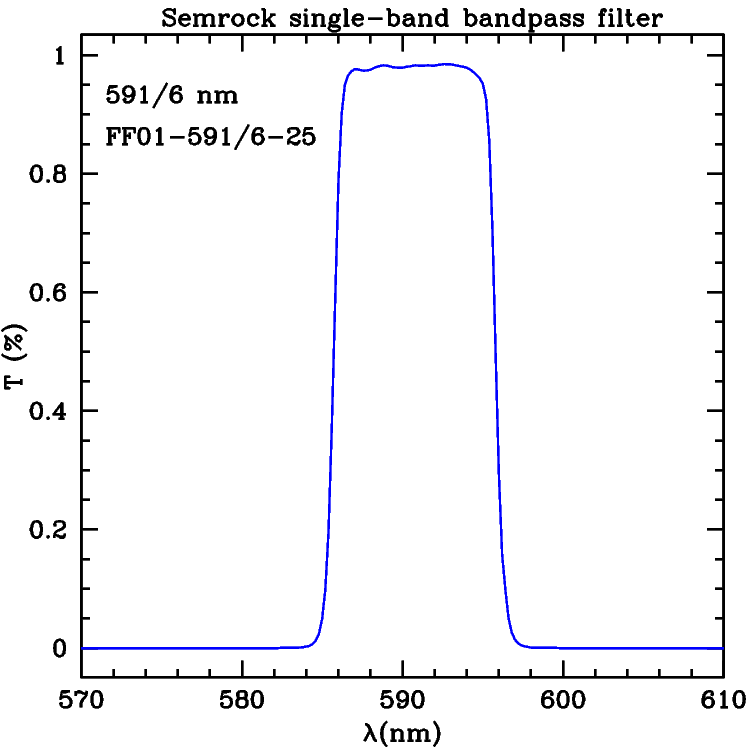}}
    \caption{Semrock bandpass filter transmissivity $(591\pm 5)\unit{nm}$.}
    \label{fig:Filter}
\end{figure}

The main characteristics of the AntarctiCor bandpass filter used during the missions are shown in Fig.~\ref{fig:Filter} and in Table~\ref{tab:AntarctiCor_features}. More information can be found in~\cite{Filter}. 

\begin{table}[ht]
    \caption{Main optical and physical filter specifications~\citep{Filter}. }
    \label{tab:AntarctiCor_ch}
    \begin{tabular}{ll}
        \hline
        \textbf{Specification} & \textbf{Value}\\
        \hline
        Transmission band                & $T_{avg} > 93\%$ at $588-594.5\unit{nm}$  \\ 
        Center wavelength                & $591.25\unit{nm}$                         \\
        FWHM bandwidth (nominal)         & $10\unit{nm}$                             \\
        Transverse dimensions (diameter) & $25\unit{mm}$                             \\
        Filter thickness                 & $5.0\unit{mm}$                            \\
        \hline
    \end{tabular}
\end{table}

Indeed, the telescope design is derived from the design of the ASPIICS space coronagraph~\citep{Galy_2015}. Some modification from the original design have been adopted due to the main difference between ASPIICS and AntarctiCor: the former is externally-occulted, the latter is internally occulted. For example, since the objective doublet lens of ASPIICS operates in the shadow of the external occulter while the AntarctiCor objective is directly illuminated by the Sun-disc light, to minimize the internal reflection in the objective lens, this has been changed into a highly polished singlet, i.e. $0.5\unit{nm}$ rms~(\citeauthor{Fineschi_2019}~\citeyear{Fineschi_2019}).

\begin{figure}[!tbp]
  \centering
  \subfloat{\includegraphics[width = 0.48\textwidth]{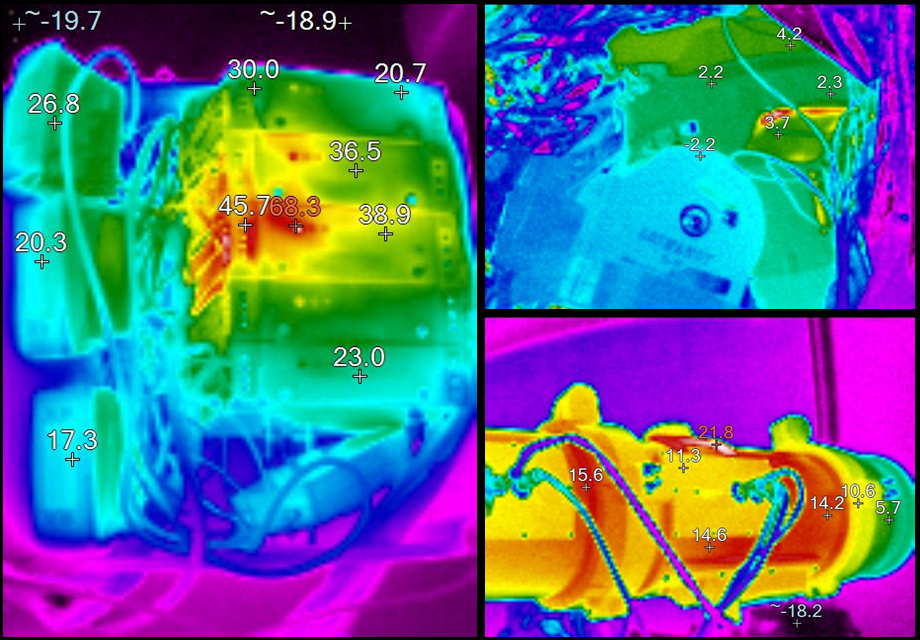}\label{fig:temperature_controll_1}}
  \hfill
  \subfloat{\includegraphics[width = 0.51\textwidth]{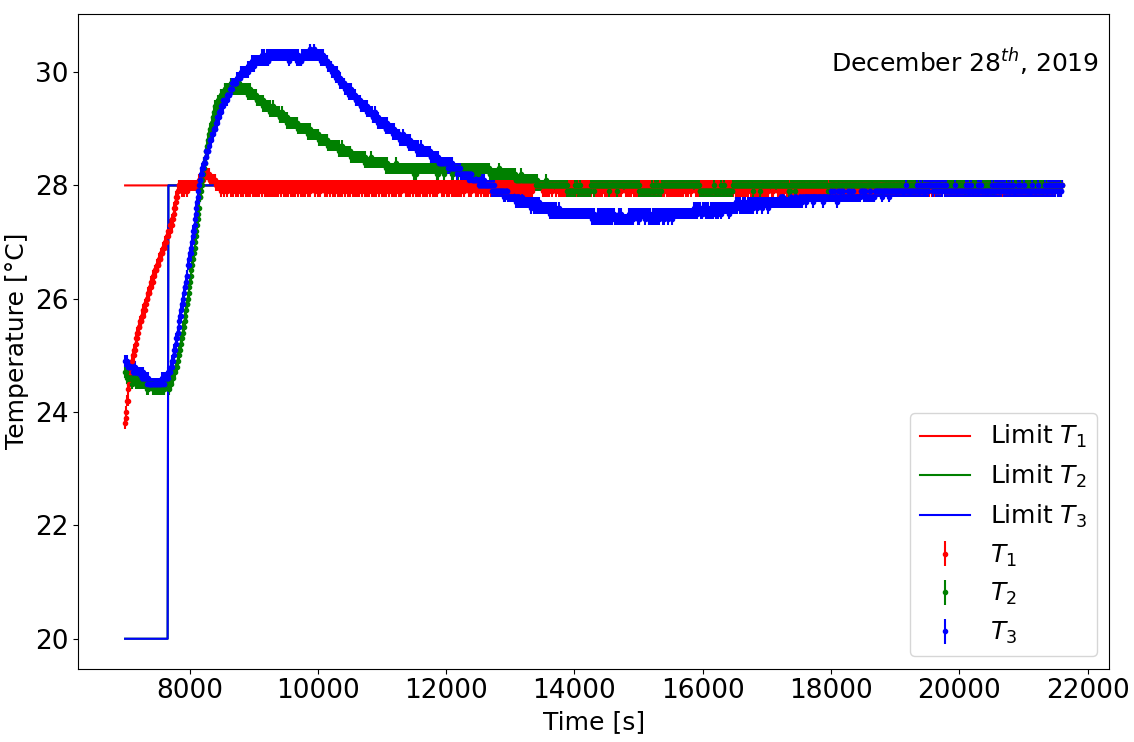}\label{fig:temperature_controll_2}}
  \caption{\textbf{Left:} Infrared camera thermal image of the instrumentation. \textbf{Right:} Example of temperature control with a limit set of 28\textdegree.}
  \label{fig:temperature_controll}
\end{figure}

The telescope temperature is acquired in three different points by three PT100 and it is controlled by using three heaters with a power of 90\unit{W}, 100\unit{W}, 40\unit{W} in the front, central, and rear subassembly, respectively. The closed-loop heater system keeps the instrument at the set temperature (Fig.~\ref{fig:temperature_controll}, right plot). The entire structure is kept at a constant temperature of~$\approx28$\textdegree. An infrared camera is used to verify the temperature of the telescope, mount, and the whole instrumentation (Fig.~\ref{fig:temperature_controll}, left panel).

\subsection{PolarCam Micropolariser Camera}
\label{ss:PolarCam_micropolarizer_camera}
\begin{figure}
    \centerline{\includegraphics[width = 1.00\textwidth, clip=]{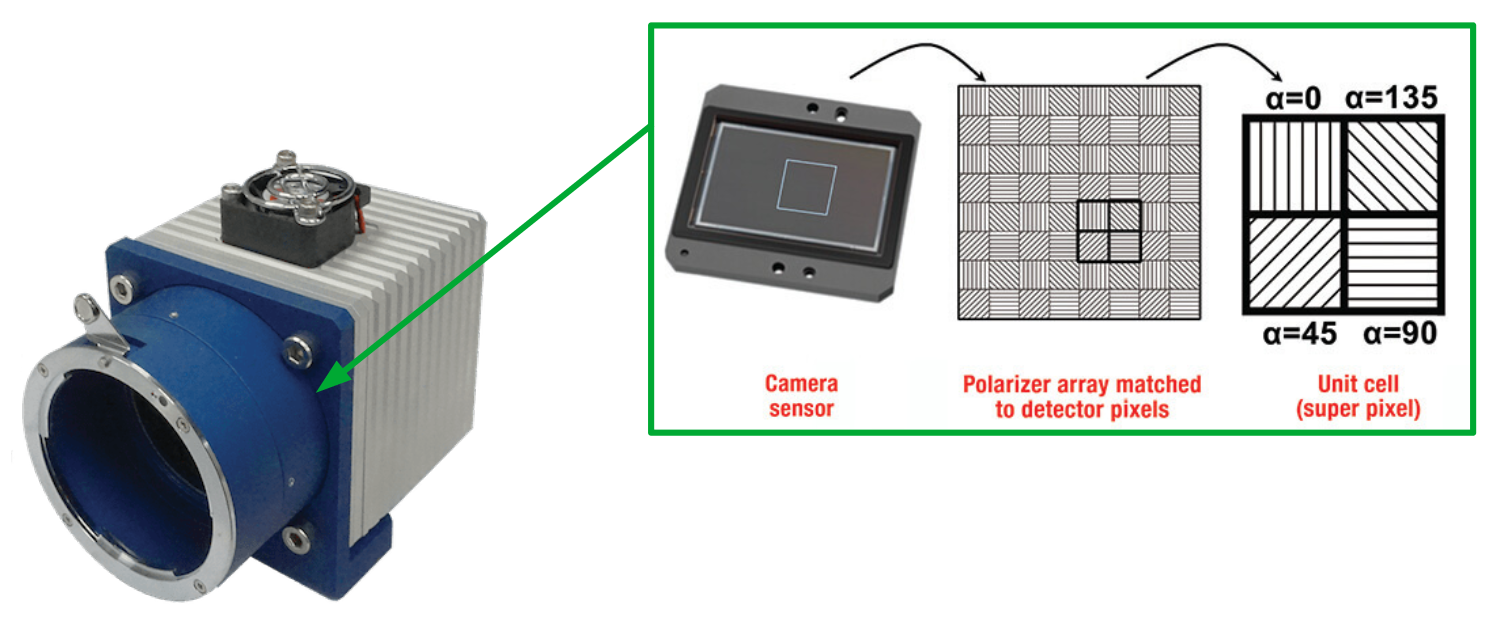}}
    \caption{PolarCam detector. An array of linear micropolarisers with different orientation are matching the pixels of the sensor~\citep{Tech4D_PolarCam}.}
    \label{fig:PolarCam}
\end{figure}

\begin{figure}[ht]
   \centerline{\includegraphics[width = 1.00\textwidth, clip=]{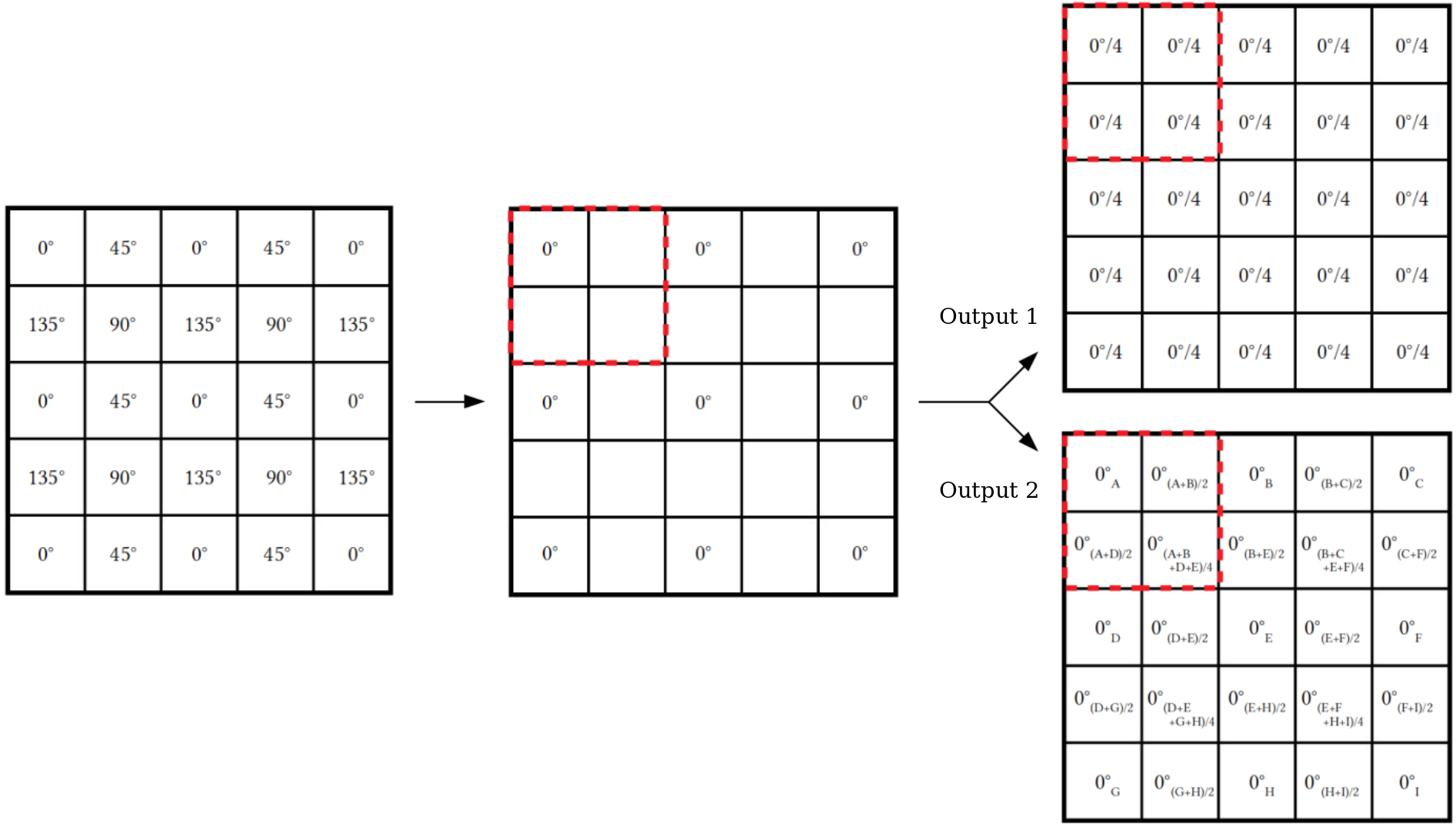}}
   \caption[example] 
   { Demosaic example to obtain a polarised image from the original raw image. In this example, the image with a polarisation angle of 0\textdegree~($I_0$) is obtained considering each super-pixel equal to the pixel value with the micropolariser at 0\textdegree~(Output 1) or considering each super-pixel obtained considering the pixel values with the micropolariser at 0\textdegree-pixels (Output 2).}
    \label{fig:PolarCam_degrid}
\end{figure}

The telescope detector is the PolarCam micropolariser camera, U4 model.\footnote{The PolarCam\textsuperscript{\textcopyright} is manufactured by 4-D Technology Corporation, Arizona, United States~\citep{4Dtech}} This camera captures simultaneously 4 images at multiple polarised angles (0\textdegree, 45\textdegree, 90\textdegree,~and 135\textdegree) thanks to an array of linear micropolarisers directly applied on the camera sensor. Different micropolariser orientations match different pixels of the sensor as shown in Fig.~\ref{fig:PolarCam}. In this way, a single acquisition can return the linear polarisation of the image as derived from the Stokes vector parameters: \textbf{S}~=~(I,~Q,~U) = ($I_0 + I_{90}$, $I_0 - I_{90}$, $I_{45} - I_{135}$) where $I_0$, $I_{90}$, $I_{45}$, $I_{135}$, are the intensities of the linear polarisation components at 0\textdegree, 90\textdegree, 45\textdegree, 135\textdegree. Indeed, the linearly polarised light requires the measurement of the I, Q, U quantities to be fully characterised~\citep{Collett_1992}. For example, it is possible to obtain the polarimetric brightness defined as $pB = \sqrt{Q^2 + U^2}$ or the sky-brightness considering the first Stokes parameter. To obtain the single $I_i$ (where $i$ = 0\textdegree, 45\textdegree, 90\textdegree, 135\textdegree) we need to perform a demosaicization process. Once a certain $I_i$ has been chosen (e.g. $I_0$), we considered the three remaining pixels of each super-pixel\footnote{A super-pixel is the \say{pixel} obtained by considering the four adjacent pixels with different micropolariser orientation (0\textdegree, 45\textdegree, 90\textdegree,~and 135\textdegree).}. The values of these single pixels are obtained as the average between the pixels (with the considered polarisation $I_0$) in each adjacent super-pixel as shown in Fig.~\ref{fig:PolarCam_degrid}, Output 2. The same procedure can be applied to get the images with the other orientations. More information about this camera and its usage can be found in~\cite{Liberatore_2021} and~\cite{Tech4D_PolarCam}. 

\section{Sky Brightness Results} 
\label{s:Results} 
During the Antarctic campaigns, systematic images of the sky were acquired in order to evaluate its brightness. 
The sky brightness in units of solar disc brightness ($B_{sky}\unit{[B_\odot]}$) was measured the first time during the 34th Campaign (austral summer 2018/2019) and during the full data acquisition during the 35th Campaign (austral summer 2019/2020) at almost regular intervals during all days.\footnote{The 36th Campaign (austral summer 2020/2021) was limited to logistical activities due to the COVID-19 pandemic.} The $B_{sky}\unit{[B_{\odot}]}$ can be evaluated by considering the first Stokes parameter $I$ and performing a ratio between what is obtained pointing to the sky~($I_{sky}$) and what is obtained pointing to the Sun using a diffuser~($I_{diff}$). Both quantities must be normalized by the respective exposure time~$t_{exp}^i$. Then, by considering the diffuser transmissivity $T_{diff} \approx28\%$ \citep{diffuser} we obtain: 

\begin{equation}
 \frac{\overline{I}_{sky}}{\overline{I}_{diff}}= \frac{I_{sky}/t_{exp}^{sky}}{I_{diff}/t_{exp}^{diff}} = \frac{B_{sky}}{(B_{\odot}\cdot T_{diff})}  
\end{equation}

\noindent Finally, by considering the light scattered over the solid angle by the diffuser over the Sun angular radius:  $\vartheta = 16'$:
\begin{equation}
    \frac{\Omega_\odot}{2\pi} = \int^{2\pi}_0\int^\vartheta_0 \frac{\sin{\vartheta'}}{2\pi}d\vartheta'd\varphi = 1-\cos{\vartheta} = 1.083\times10^{-5} 
\end{equation}
the resulting sky brightness $B_{sky}\unit{[B_\odot]}$ is~\citep{OpalFormula}:

\begin{equation}
    B_{sky}[B_{\odot}] = \frac{(I_{sky}/t_{exp}^{sky})}{(I_{diff}/t_{exp}^{diff})}\cdot T_{diff}\cdot \frac{\Omega_\odot}{2\pi}
\end{equation}

\begin{figure} 
    \centerline{\includegraphics[width = 0.70\textwidth, clip=]{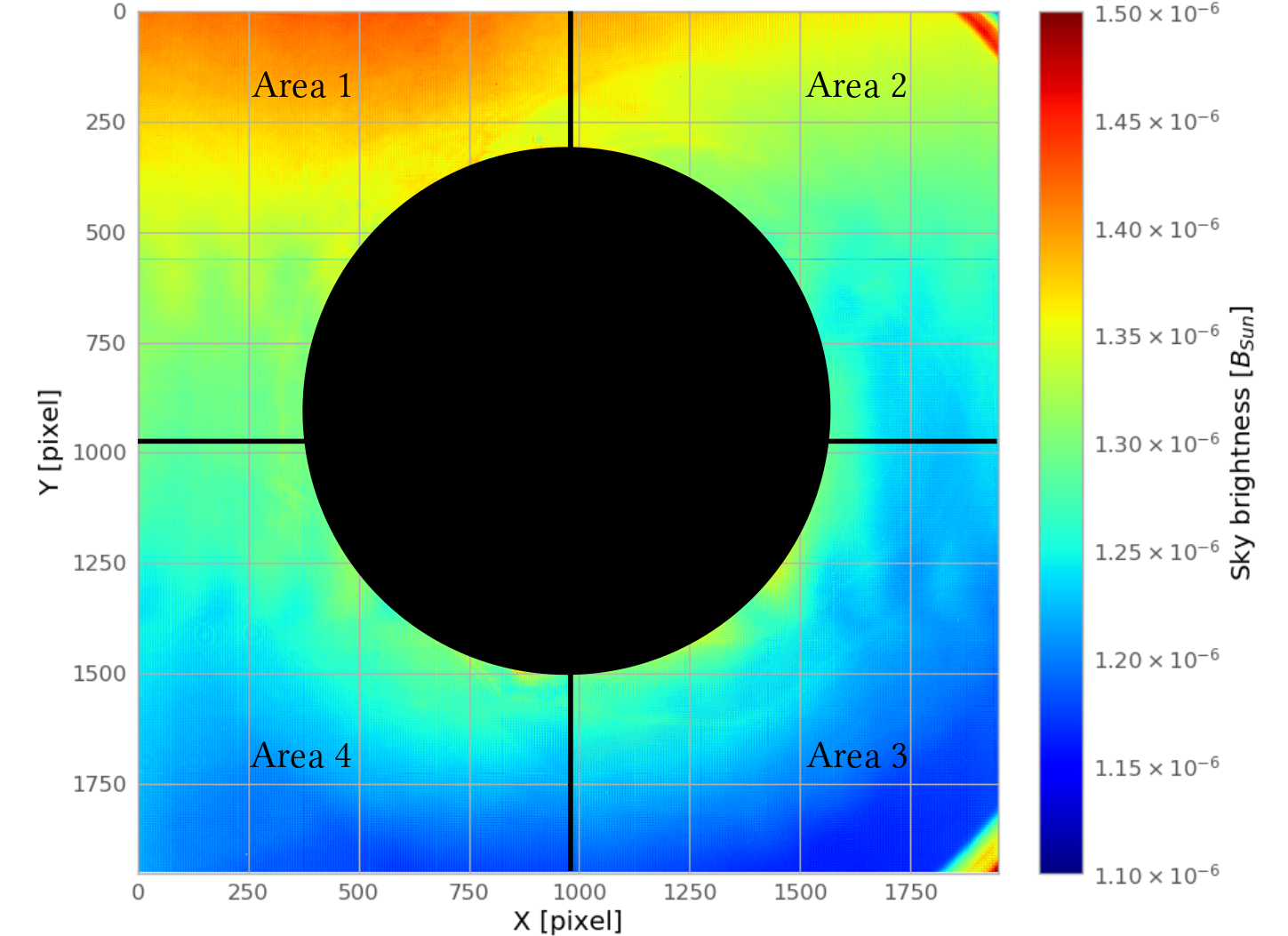}}
    \caption{Example of the measured sky brightness (Dome C, Concordia Station - Antarctica) from the 34th campaign. The entire frame is divided into 4 different pads. The final sky brightness is obtained bye averaging the 4 areas.}
    \label{fig:frame_4division}
\end{figure}

In particular, in our case, the acquired $B_{sky}[B_{\odot}]$ frame was divided into four different regions (Area in Fig.~\ref{fig:frame_4division}) and the final brightness was obtained by averaging them:
\begin{equation}
    B_{sky}[B_{\odot}] = \frac{\sum_i B_{sky}^i}{4} \qquad [i = 1, 2, 3, 4].
\end{equation}

\noindent For each $B_{sky}$ we evaluate the dispersion values as the standard deviation over the pixels in a considered area (e.g. $\sigma$ over Area 1) quadratically added (and then divided by a factor of four) to the standard deviations obtained in the other three regions (Area 2, 3, 4).

A so-called \say{pure blue sky} (i.e. $B_{sky}\approx10^6\unit{B_{\odot}}$) is necessary to carry out ground-based observations of the corona~\citep{fracast, Fracast_Righini, Elmore}.
\\
Figure~\ref{fig:brightness_solar_corona_components} shows the sky brightness, in $\unit{B_{\odot}}$ units, measurements obtained during the 35th Italian Antarctic Expedition (2019-2020), at Concordia Station - Dome C. 

From the images acquired during the 34th Campaign on January 08, 2019 (pointing at RA: 22\textdegree~42' 32.3'', Dec: -22\textdegree~ 11' 22'') and January 09, 2019 (pointing at RA: 22\textdegree~38' 33.3'', Dec: -22\textdegree~18' 19''), we obtain that:
\begin{equation}
    B_{sky}= (1.2 \pm 0.1)\times10^{-6} B_{\odot}.
\end{equation}

\begin{figure} 
    \centerline{\includegraphics[width = 1.00\textwidth, clip=]{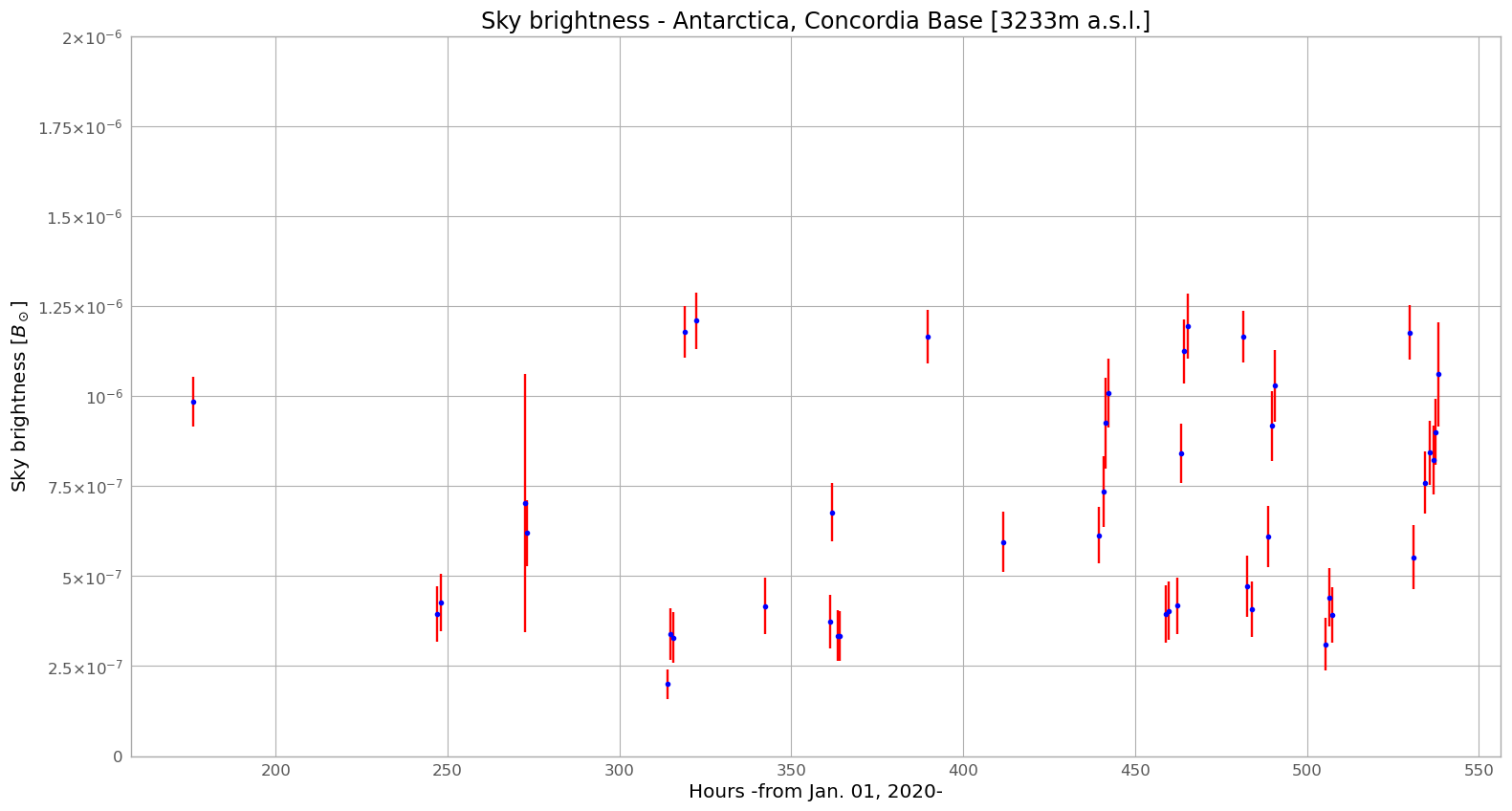}}
    \caption{Sky brightness$\unit{[B_{\odot}]}$ measurements obtained during the 35th Italian Antarctic Campaign (2019-2020), Concordia Station - Dome C, $\approx3230\unit{m}$ a.s.l., Antarctica. On the x-axis, the acquisition UTC time from January 1$^{st}$, 2020 to January 22$^{nd}$, 2020. The bars represent the dispersion values obtained by considering the standard deviations for each of the four different detector frame areas.}
    \label{fig:SkyBrightness_Bsun}
\end{figure}

During the 35th campaign it was possible to perform more systematic sky-brightness measurements. A summary of what obtained is shown in Fig.~\ref{fig:SkyBrightness_Bsun}. Averaging over the different values we obtain: 
\begin{equation}
\bar{B}_{sky} = (6.9 \pm 0.2) \times 10^{-7} \unit{B_\odot}.
\end{equation}
During these measurements, Dome C showed the characteristics of a \say{coronagraphic sky}. Sometimes, the presence of clouds, high wind or excessive suspended ice in the atmosphere made it impossible to perform good observations. On the other hand, during this campaign, we evaluated a percentage of good-weather days around the 80\%!

\begin{figure} 
    \centerline{\includegraphics[width = 1.00\textwidth, clip=]{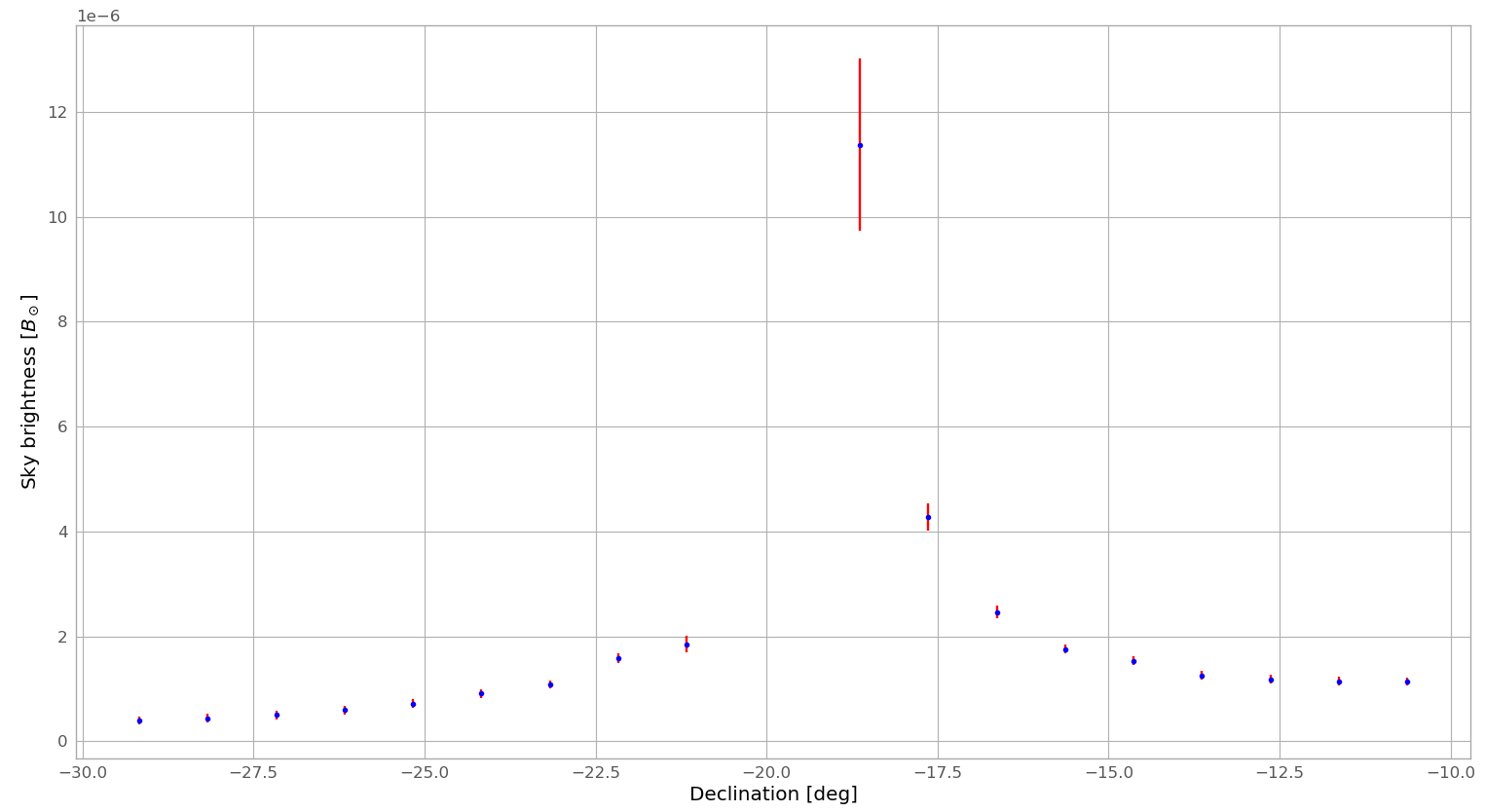}}
    \centerline{\includegraphics[width = 1.00\textwidth, clip=]{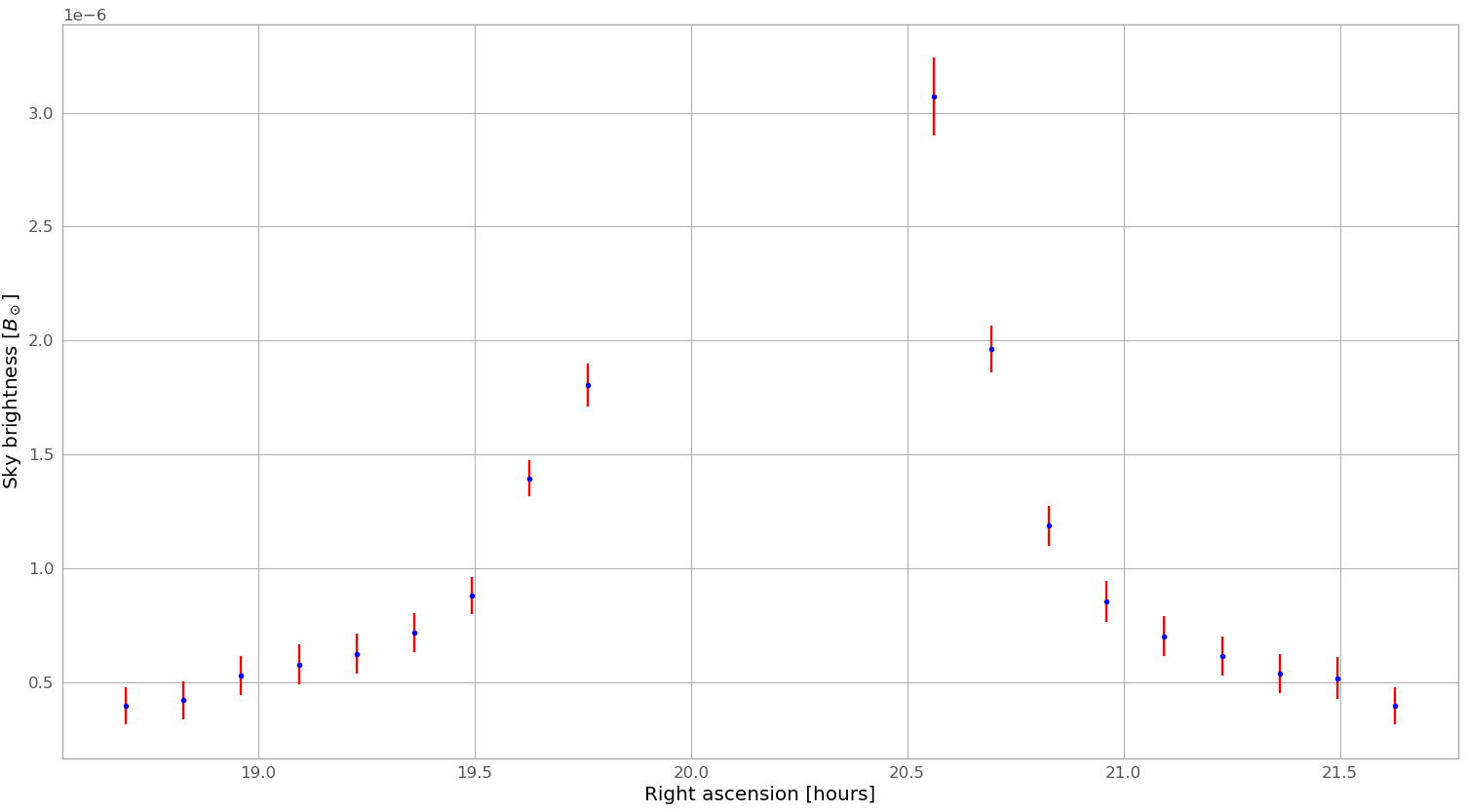}}
    \caption{Sky brightness$\unit{[B_{\odot}]}$ measurements obtained during the 35th Italian Antarctic Campaign (2019-2020), Concordia Station, Dome C, $\approx3230\unit{m}$ a.s.l., Antarctica, for different declination values (up) and different right ascension ones (down) on January 1, 2020. The Sun position (J200 system) was AR: 20\textdegree~09$^\prime$ 38$^\prime$$^\prime$, Dec: -20\textdegree~05$^\prime$ 57$^\prime$$^\prime$. }
    \label{fig:SkyBrightness_Bsun_CROX}
\end{figure}

During the same mission, we evaluated the sky brightness for different right ascensions (RA) and declinations (DEC) for a fixed day (see Fig.~\ref{fig:SkyBrightness_Bsun_CROX}). The values of measured $B_{sky}$ are consistent with what obtained in Fig.~\ref{fig:SkyBrightness_Bsun} and, as expected, it is possible to see a decrease of sky-brightness moving away from the Sun. The closest measurements to the solar limb are at about 0.7\textdegree, e.g. for the declination, the closest measurement was at 1\textdegree~from Sun center (Fig.~\ref{fig:SkyBrightness_Bsun_CROX}, up), so $1 - 0.26 = 0.74$\textdegree~from the solar limb (where 0.26\textdegree~is the solar radii). We cannot really go much closer than 1\textdegree~from Sun center due to the instrument field of view along the $x$ and $y$ axis. The instrument FoV is $\pm0.6$\textdegree~(Table~\ref{tab:AntarctiCor_features}) and we need that the Sun is completely out to take sky brightness measuraments.\\

These results show that the Antarctic sky at Dome C, where Concordia Station is located, has periods when it can be considered to have a \say{coronagraphic sky} (e.g.~\citeauthor{fracast}~\citeyear{fracast};~\citeauthor{Fracast_Righini}~\citeyear{Fracast_Righini};~\citeauthor{Elmore}~\citeyear{Elmore} and  Fig.~\ref{fig:brightness_solar_corona_components}).

\section{Conclusion} 
\label{s:Conclusion} 
In the present work, we presented the first results from the ESCAPE project. In particular, after a description of the project and its main goal, we described the instrumentation used during the Antarctic missions and its characteristics. We described the internally-occulted coronagraph AntarctiCor and its innovative detector with arrays of micropolariser for linear-polarisation imaging. Then, we described how we evaluated the sky brightness $\bar{B}_{sky}$ at Concordia Station (Dome~C plateau - Antarctica, coord. 75\textdegree~06$^\prime$~S; 123\textdegree~20$^\prime$~E) at an altitude of $\approx3300\unit{m}$ above sea level. The $\bar{B}_{sky}$ value was obtained during two different missions. In the first one (34th Italian expedition to Antarctica - austral summer 2018/19), a sky brightness of $\approx 1.0 \times 10^{-6} \unit{B_\odot}$ was measured. Due to a logistic problem, just few acquisitions during the mission were possible (from 18 January 2019 to 19 January 2019). These sky brightness values have been obtained for a fixed distance from the Sun. subsequently, during the 35th mission, we performed more systematic measurements (from 1 January 2020 to 22 January 2020). In addition, we performed measurements of the sky brightness not only for a fixed distance from the Sun, but for different declination and different right ascension values. In this second campaign we obtained that $\bar{B}_{sky} \approx 7.0 \times 10^{-7} \unit{B_\odot}$. Both results quantitatively demonstrate, for the first time, the quality of the Dome~C site for coronagraphic observations. We can conclude that the Antarctic sky at Concordia Station shows the characteristics of a \say{coronagraphic sky} (i.e., $B_{sky} < 10^6\unit{B_{\odot}}$). This holds the promises for Concordia Station to host a permanent coronagraph observatory for continuous studies of the solar corona, during the Austral summers.


\begin{acks}
This paper has been possible thanks to the whole INAF and ESCAPE Project team and the Italian Piano Nazionale Ricerche Antartico~\citep{PNRA}. The AntarctiCor - ESCAPE project is funded by the PNRA, grant N. 2015-AC3.02. The authors thank Dr. Angelo Lupi (PI of BSRN project) for providing us with access to the images of the AstroConcordia all-sky camera. A particular acknowledgement to OPTEC S.p.A for the AntarctiCor telescope thermal and structural design and realisation as well. The authors thank also ALTEC Company for providing logistic support during the many AntarctiCor calibration periods at the INAF OPSys facility. PROBA-3/ASPIICS is an ESA mission.\\\\
\end{acks}

\begin{conflict}
The authors declare that there are no conflicts of interest.
\end{conflict}

\begin{dataavailability}
The datasets generated during and/or analysed during the current study are available from the corresponding author on reasonable request.
\end{dataavailability}





\bibliographystyle{spr-mp-sola}
\bibliography{Manuscript.bib} 

\end{article} 
\end{document}